\documentclass[twocolumn,superscriptaddress,showpacs]{revtex4-2}  % for review and %submission
\PassOptionsToPackage{hyphens}{url}\usepackage{hyperref}
\usepackage{hyperref}
\usepackage{breakurl}
\usepackage{amssymb,amsmath,amsfonts}   % for math
\usepackage{tikz} 
\usepackage{graphicx}  % needed for figures
\usepackage{chngcntr}
\usepackage{apptools}
\usetikzlibrary{decorations.markings}

\usepackage{bibentry}
\usepackage{natbib}
\usepackage{graphicx}  % needed for figures
\usepackage{dcolumn}   % needed for some tables
\usepackage{bm}        % for math
\usepackage{empheq}
\usepackage{enumerate}
\usepackage{slashed}
\usepackage{simplewick}
\usepackage{comment}
\usepackage{color}
\usepackage{soul}
\usepackage[english]{babel}
\usepackage{appendix}
\usepackage[utf8]{inputenc}

\usepackage{marginnote}
\usepackage[normalem]{ulem}

\def\Pic{\mbox{Pic}}
\newcommand{\eq}[1]{(\ref{#1})}

\newcommand{\ba}{\begin{align}}
\newcommand{\ee}{\end{equation}}
\newcommand{\be}{\begin{equation}}

\def\12{\frac{1}{2}}
\def\3{\frac{1}{2\pi}}
\newcommand{\p}{\partial}
\newcommand{\en}{\end{align}}

\usepackage{color}

\def\g{{\rm g}}

\begin{document}

\setcounter{secnumdepth}{-1}

\title{Geometric test for topological states of matter}
\author{S. Klevtsov}
\affiliation{IRMA, Universit\'e de Strasbourg, UMR 7501, 7 rue Ren\'e Descartes, 67084 Strasbourg, France}
\author{ D. Zvonkine}
 \affiliation{ CNRS, Universit\'e de Versailles St-Quentin (Paris-Saclay), Versailles, France}%Lines 

%  automatically or can be forced with \\
%  \email{Second.Author@institution.edu}
% \affiliation{%
%  \homepage{http://www.Second.institution.edu/~Charlie.Author}
% \affiliation{


\begin{abstract}
We generalize the flux insertion argument due to Laughlin, Niu-Thouless-Tao-Wu, and Avron-Seiler-Zograf to the case of fractional quantum Hall states on a higher-genus surface. We propose this setting as a test to characterise the robustness, or topologicity, of the quantum state of matter and apply our test to the Laughlin states.  

Laughlin states form a vector bundle, the Laughlin bundle, over the Jacobian -- the space of Aharonov-Bohm fluxes through the holes of the surface. The rank of the Laughlin bundle is the degeneracy of Laughlin states or, in presence of quasiholes, the dimension of the corresponding full many-body Hilbert space; its slope, which is the first Chern class divided by the rank, is the Hall conductance. We compute the rank and all the Chern classes of Laughlin bundles for any genus and any number of quasiholes, settling, in particular, the Wen-Niu conjecture. Then we show that Laughlin bundles with non-localized quasiholes are not projectively flat and that the Hall current is precisely quantized only for the states with localized quasiholes. Hence our test distinguishes these states from the full many-body Hilbert space.

\end{abstract}
\date{\today}
\maketitle
%\nocitenames
\noindent{\it 1. Introduction.}\; Topological states describe special phases of strongly-correlated quantum matter arising at low temperatures, which exhibit certain remarkable properties, such as precise quantization phenomena in materials with impurities, fractional and non-abelian statistics, and ground state degeneracy robust under local perturbations. These unusual properties make the topological states suitable for a range of applications from quantum metrology to fault-tolerant quantum computing. 

In this paper we ask the following question: given the ground state of a quantum system, how can we determine if it describes a topological state of matter? What kind of a criterion can be given to characterize the robustness of topological states? Here we give a concrete criterion, a geometric test that identifies topological states. Our criterion is of geometric nature and applies to situations where the ground state of the system is degenerate.

Best known examples of topological states of matter include superconductors, spin liquids, quantum Hall states etc. We focus here on the fractional quantum Hall effect, where explicitly defined trial states are available for analytic investigation, although the criterion is potentially applicable to a broader class of quantum phases of matter. 

\vspace{0.3cm}
\noindent{\it 2. Projective flatness test for the ground state bundle.} \;Here we describe our test in a general setting. We consider the situation of a non-abelian Berry connection, see Ref.\ \cite{WZ1984}, when the ground state is degenerate and separated by a gap from the rest of the spectrum, and depends continuously on an $n$-dimensional manifold $M$ of classical parameters. We further place ourselves in the framework of the adiabatic theorem where the degeneracy of the ground state is constant over the whole parameter manifold. Our ground states thus form a hermitian vector bundle $V$ of quantum states over the parameter space~$M$. The rank $r$ of~$V$ is the degeneracy of the ground state. In this general situation, the vector bundle has a set of Chern classes $c_1(V),c_2(V),c_3(V)$, terminating at $c_m$ for $m=\min\{[\frac n2],r\}$, where $[\cdot]$ denotes the integral part. 
The rank $r$ in general can be a function of the number of particles, flux of the magnetic field, number of quasiparticles, geometry and topology of the surface as well as any other parameters of the problem. 

Topological states are further characterized by their robustness against perturbations. We formalize this condition for our vector bundle as the requirement that the adiabatic transport of the quantum states along a path $\gamma$ in the parameter space  $M$ is independent of continuous deformations of the path and depends only on its topology, possibly up to a $U(1)$ Berry phase. In more precise terms, we require that the adiabatic transport defines a projectively flat connection on~$V$. A connection like that induces a representation $\rho:\pi_1\to {\rm PGL}(r,\mathbb C)$, where $\pi_1$ is the first fundamental group of $M$ and ${\rm PGL}(r,\mathbb C)={\rm GL}(r,\mathbb C)/{\mathbb C}^*$ is the projective linear group (factoring by $\mathbb C^*$ allows for the nontrivial $U(1)$ phase). 

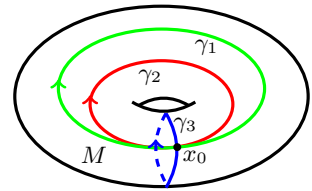
\begin{figure}[h]
\begin{center}
\begin{tikzpicture}[smooth cycle,scale=0.30]
\tikzset{->-/.style={decoration={
  markings,
  mark=at position .5 with {\arrow{>}}},postaction={decorate}}}
  \tikzset{-<-/.style={decoration={
  markings,
  mark=at position .5 with {\arrow{<}}},postaction={decorate}}}
\draw [black,very thick] (6,6.35) ellipse (6.5cm and 4cm);

\draw [red,very thick,bend left,-<-] (3.5+2.5,3.5+2.5) ellipse (90pt and 55pt);
\draw [green,very thick,bend left,-<-] (3.5+2.5,3.5+3.2) ellipse (130pt and 75pt);
\draw [blue,very thick,bend right]  (3.5+2.7,-.2+2.5) edge [-] (3.5+2.7,3.1+2.5);
\draw [blue,dashed,very thick,bend right]  (3.5+2.7,3.3+2.3) edge [-<-] (3.5+2.7,0.2+2.3);
\draw[black] (7.5,3) node [above]{$x_0$};
\draw[black] (8,7.8) node [above]{$\gamma_1$};
\draw[black] (5.5,6.5) node [above]{$\gamma_2$};
\draw[black] (7.05,4.35) node [above]{$\gamma_3$};
\draw[black] (3,3) node [above]{$M$};
\fill [black] (6.7,4.1) circle (5pt);
\draw [black,very thick,bend right] (11-6.3,3.7+2.4) edge (13.8-6.3,3.7+2.4);
\draw [black,very thick,bend left] (11.3-6.3,3.55+2.4) edge (13.5-6.3,3.55+2.4);
%\draw [black,dashed,very thick,bend right] (10.01+9,0.99) edge (5.96+9,.96);
%\draw [black,dashed,very thick,bend right] (6.1+9,6.35) edge (10.04+9,6.39);
%\draw [black,very thick,bend right] (2,3.7) edge (4.8,3.7);
%\draw [black,very thick,bend left] (2.3,3.55) edge (4.5,3.55);
%\draw [black,very thick,bend right] (11+9,3.7) edge (13.8+9,3.7);
%\draw [black,very thick,bend left] (11.3+9,3.55) edge (13.5+9,3.55);
%\draw [red,dashed,very thick,bend right,->-] (5.7,3.8) arc (-20:-160:2.4);
%\draw [green,dashed,very thick,bend right,->-] (5.7+9,3.8) arc (-20:-160:2.4);
%\draw [violet,dashed,very thick,bend right,->-] (5.7+18,3.8) arc (-20:-160:2.4);
%\draw [blue,very thick]  (3.5,3.5) edge [->] (3.5,6);
%%\draw [blue,dashed,very thick,bend right]  (3.5,3.3) edge [-<-] (3.5,0.2);
%\draw [cyan,very thick]  (3.5+9,3.5) edge [->] (3.5+9,6);
%%\draw [cyan,dashed,very thick,bend right]  (3.5+9,3.3) edge [-<-] (3.5+9,0.2);
%\draw [orange,very thick]  (3.5+18,3.5) edge [->] (3.5+18,6);
%\draw[black] (4.6,4.6) node [above]{$\phi_1$};
%\draw[black] (4.2,.4) node [above]{$\phi_{1+\g}$};
%\draw[black] (4.6+9,4.6) node [above]{$\phi_2$};
%\draw[black] (4.2+9,.4) node [above]{$\phi_{2+\g}$};
%\draw[black] (4.6+18,4.6) node [above]{$\phi_\g$};
%\draw[black] (4.2+18,.4) node [above]{$\phi_{2\g}$};
% %       \draw[black] (8.15,3.55) node [above]{$P_0$};
%  %              \fill [black] (8,3.6) circle (3pt);
\end{tikzpicture}
{\small \caption{The adiabatic transport of quantum states 
in the parameter space $M$ along the curves $\gamma_{1},\gamma_2,\gamma_3$ starting and ending at the same point $x_0$.}
\label{fig:moduli}}
\end{center}
\end{figure}

For example, in Fig.\ \ref{fig:moduli}, the adiabatic transport along the curves $\gamma_1$ and $\gamma_2$, which can be continuously deformed into each other, would have the holonomies equivalent up to $U(1)$ phase, while the transport along $\gamma_3$ could yield an apriori different holonomy.

By a standard result on complex vector bundles \cite[Cor.\ 2.7]{Kob2014}, if $V$ is projectively flat then its total Chern class, which is the sum of all Chern classes $c(V)=1+c_1(V)+c_2(V)+\cdots+c_n(V)$ is given by 
\[
c(V)=\left(1+\frac{c_1(V)}r\right)^r.
\]
In other words, the higher Chern classes of a projectively flat bundle~$V$ are given by powers of its first Chern class 
\begin{equation}\label{test}
c_i(V)={{r}\choose{ i}}\cdot \left[\frac{c_1(V)}r\right]^i,\quad1\leq i\leq \min\left\{\left[\frac n2\right],r\right\},
\end{equation}
times the binomial coefficient.
No such relations exist for a general vector bundle, and if the relations do not hold the quantum state cannot be topological in the sense above.

If $r=1$, the ground state is non-degenerate and the vector bundle becomes a line bundle.  Hence it does not have higher Chern classes. On the other hand, parameter spaces of the real dimension $\dim M\leq3$ do not support higher Chern classes either. Therefore the projective flatness test is only applicable when $r>1$ and $\dim M\geq4$. This brings us naturally to the framework of Laughlin states on surfaces of genus $\g \geq 2$.

\vspace{0.3cm}
\noindent{\it 3. Laughlin states on Riemann surfaces.} \; A Laughlin trial state, Ref.\ \cite{Laughlin1983}, describes the fractional quantum Hall effect for filling fractions of the form $1/\beta$, where $\beta$ is an odd integer. Here we define an $N$-particle Laughlin state with $p$ quasiholes on the Riemann sphere for any positive integer $\beta$ as follows:
\begin{equation}\label{Ls}
\Psi_L=P(z_1,z_2,\ldots,z_N)\prod_{1 \leq n<m \leq N} (z_n-z_m)^\beta;
\end{equation}
where $P$ is a completely symmetric polynomial in $N$ variables $z_1,...,z_N\in\mathbb C$, of degree at most~$p$ in each~$z_n$. Thus we do not restrict to the fermionic case and consider the bosonic states as well. Besides the (anti-) symmetry for (odd) even $\beta$, the two other defining properties of $\Psi_L$ is the vanishing on the diagonal $\Delta=\cup_{n<m}\{z_n=z_m\}$ to the order $\beta$ and total degree in each variable $z_n$ being equal to the magnetic flux
\[N_\phi=p+\beta(N-1).\]
The vanishing on the diagonal takes into account the Coulomb interactions between the particles and the total degree ensures that each electron is on the lowest Landau level for the magnetic field with flux~$N_\phi$.

The full many-body Hilbert space of functions of the form~\eq{Ls} has dimension ${N+p}\choose{p}$, by the number of linearly independent polynomials~$P$. 
It has a special one-dimensional subspace of states with $p$ quasiholes localized at positions $w_1,w_2,\ldots,w_p$:
\begin{equation}\label{Lsq}
\Psi_L=\prod_{i=1}^p\prod_{n=1}^N(z_n-w_i)\prod_{n<m}^N(z_n-z_m)^\beta.
\end{equation}
As long as the positions of the quasiholes are fixed, the states above do not depend on any continuous parameters. 

\begin{figure}[h]
\begin{center}
\vspace{.5cm}
\begin{tikzpicture}[smooth cycle,scale=0.30]
\tikzset{->-/.style={decoration={
  markings,
  mark=at position .5 with {\arrow{>}}},postaction={decorate}}}
  \tikzset{-<-/.style={decoration={
  markings,
  mark=at position .5 with {\arrow{<}}},postaction={decorate}}}
\draw [black,very thick,bend right] (6,6.35) arc (50:310:3.5);
\draw [black,very thick,bend right] (10+9,1) arc (-130:130:3.5);
\draw [black,very thick,bend right] (10.01,0.99) edge (5.96,.96);
\draw [black,very thick,bend right] (6,6.35) edge (10.04,6.39);
\draw [black,very thick,bend left] (10+5,0.97) edge (6.0+4,.99);
\draw [black,very thick,bend left] (6+4,6.35) edge (10.04+5,6.39);
\draw [black,very thick,bend right] (11,3.7) edge (13.8,3.7);
\draw [black,very thick,bend left] (11.3,3.55) edge (13.5,3.55);
\draw [black,dashed,very thick,bend right] (10.01+9,0.99) edge (5.96+9,.96);
\draw [black,dashed,very thick,bend right] (6.1+9,6.35) edge (10.04+9,6.39);
\draw [black,very thick,bend right] (2,3.7) edge (4.8,3.7);
\draw [black,very thick,bend left] (2.3,3.55) edge (4.5,3.55);
\draw [black,very thick,bend right] (11+9,3.7) edge (13.8+9,3.7);
\draw [black,very thick,bend left] (11.3+9,3.55) edge (13.5+9,3.55);
\draw [red,dashed,very thick,bend right,->-] (5.7,3.8) arc (-20:-160:2.4);
\draw [green,dashed,very thick,bend right,->-] (5.7+9,3.8) arc (-20:-160:2.4);
\draw [violet,dashed,very thick,bend right,->-] (5.7+18,3.8) arc (-20:-160:2.4);
\draw [blue,very thick]  (3.5,3.5) edge [->] (3.5,6);
%\draw [blue,dashed,very thick,bend right]  (3.5,3.3) edge [-<-] (3.5,0.2);
\draw [cyan,very thick]  (3.5+9,3.5) edge [->] (3.5+9,6);
%\draw [cyan,dashed,very thick,bend right]  (3.5+9,3.3) edge [-<-] (3.5+9,0.2);
\draw [orange,very thick]  (3.5+18,3.5) edge [->] (3.5+18,6);
\draw[black] (4.6,4.6) node [above]{$\phi_1$};
\draw[black] (4.2,.4) node [above]{$\phi_{1+\g}$};
\draw[black] (4.6+9,4.6) node [above]{$\phi_2$};
\draw[black] (4.2+9,.4) node [above]{$\phi_{2+\g}$};
\draw[black] (4.6+18,4.6) node [above]{$\phi_\g$};
\draw[black] (4.2+18,.4) node [above]{$\phi_{2\g}$};
 %       \draw[black] (8.15,3.55) node [above]{$P_0$};
  %              \fill [black] (8,3.6) circle (3pt);
\end{tikzpicture}
{\small \caption{AB phases on the genus-$\g$ Riemann surface.}
\label{fig:highergenus}}
\end{center}
\end{figure}
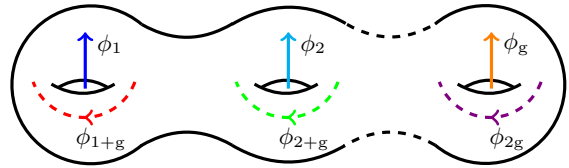

The standard way to bring a parameter space into the game in QHE is to consider Laughlin states on a Riemann surface $\Sigma$ of genus $\g>0$ \cite{HaldaneRezayi,Thouless1985,TaoWu,Avron85,Avron1994}. The definition here mimics the one given above for the sphere \cite{Kl2019}. Namely, we require the (anti-)symmetry for (odd) even $\beta$ and vanishing on the diagonal $\Delta=\cup_{n<m}\{z_n=z_m\}$ to the order~$\beta$. The analog of being a degree-$N_\phi$ polynomial on a compact Riemann surface is the condition that $\Psi_L$ is a section of a degree-$N_\phi$ holomorphic line bundle~$L$.  Now, the latter come with a natural parameter space: the moduli space of degree-$N_\phi$ line bundles is the Picard variety $\Pic^{N_\phi}(\Sigma)$ isomorphic to a $\g$-dimensional complex torus. These inequivalent configurations of the magnetic field of flux $N_\phi$ through the surface are obtained by applying the Aharonov-Bohm solenoid fluxes through the $2\g$ cycles on the surface, $\{\phi_a\}_{a=1,\ldots,2\g}\in[0,2\pi]^{2\g}$, see Fig. \ref{fig:highergenus}.

This is precisely the setting of a higher genus surface considered in the non-degenerate (IQHE) case in Ref.~\cite{Avron1994}, and generalizing Laughlin's gauge argument~\cite{Laughlin1981} (for the case of a torus see Refs.~\cite{Thouless1985,TaoWu,Avron85}). Following the standard argument of these Refs., when changing the flux through the cycle $b$ of the surface, $\phi_b=-V_bt$ adiabatically with time~$t$, the Hall current through the cycle $a$ equals $V_b$ times the Hall conductance, 
\[
I_a=(\sigma_{H})_{ab} V_b,
\]
which is the first Chern class of the Laughlin bundle over $\Pic^{N_\phi}$ divided by its rank in case of degeneracy,
\begin{equation}\label{cond}
\sigma_H=\sum_{a,b} (\sigma_{H})_{ab}\;d\phi_a\wedge d\phi_b=\frac1{{\rm rk}\, V}c_1(V).
\end{equation}
Thus we need to compute the rank and the first Chern class of the Laughlin bundle.
 
\vspace{0.3cm}
\noindent{\it 4. Quantum optimal packing problem.} \;
We begin by computing the rank of the Laughlin bundle, i.e. the degeneracy of states \eq{Lsq} on a genus-$\g$ Riemann surface. The space of sections of a holomorphic line bundle~$S$ over a complex manifold~$X$ is denoted by $H^0(X,S)$ and its dimension by $h^0(X,L)$. The main tool for computing $h^0(X,S)$ is the Hirzebruch-Riemann-Roch formula
\begin{equation} \nonumber
\sum_{i=0}^{\dim_{\mathbb C}X} (-1)^i h^i(X,S)=\int_X e^{c_1(S)}{\rm td}(X),
\end{equation}
where $h^i(X,S)$ are the dimensions of cohomology groups of~$S$, $c_1(S)$ is the first Chern class of $S$, and ${\rm td}(X)$ is the Todd class of $X$.
In the case of Laughlin states, one can use the Kodaira vanishing theorem to show that the higher cohomology groups in $H^i(X,S),\;i>0$ vanish and what remains in the left hand side is just $h^0(L,X)$, which is the degeneracy of ground states:
\begin{equation}\label{rank1}
r=\int_X e^{c_1(S)}{\rm td}(X).
\end{equation}
We do not recall the general definition of the Todd class here, but an expression for ${\rm td}(X)$ in the case of Laughlin states is given below in Eq.~\eqref{td(X)}. Let us describe $X$ and $S$ in this case. 

As a multi-particle wave function, the Laughlin state is naturally a function on the cartesian product of $N$ copies of the Riemann surface $\Sigma^N$. More precisely it is a section of the line bundle $L^{\boxtimes N}=\pi^*_1\Sigma\otimes\cdots\otimes\pi^*_NL$ over $\Sigma^N$, which is (anti-)symmetric for (odd) even $\beta$ and vanishes on the diagonal to the order $\beta$. Twisting $L^{\boxtimes N}$ by the divisor $\beta$ times the diagonal $\Delta=\cup_{n<m}^N\{z_n=z_m\}$ we reinterpret a Laughlin state as a completely symmetric section of the bundle $S = L^{\boxtimes N}(-\beta\Delta)$ over the $N$th symmetric power of the Riemann surface $X = S^N\Sigma = \Sigma^N/S_N$. Note that $S^N\Sigma$ is a smooth complex manifold: indeed, locally unordered sets $\{ z_1, \dots, z_N \}$ of $N$ complex numbers are parametrized by the coefficients of the polynomial $(z-z_1) \cdots (z-z_N)$. 

For $N\geq 2\g-1$, there is a particularly useful representation of the $N$th symmetric power $X = S^N\Sigma$ of a Riemann surface as a holomorphic bundle of projective spaces $\mathbb P^{N-\g}$ over the Picard group $\Pic^N(\Sigma)$ of the surface, see \cite[\S 3.a]{Gunning2015}. The Picard group is a $\g$-dimensional complex torus isomorphic to the Jacobian of $\Sigma$, but the isomorphism is not canonical. We use $\Pic^N$ rather than the Jacobian because it makes the map $X \to \Pic^N(\Sigma)$ canonical and also because it allows us to distinguish $\Pic^N$ from {\em another} complex torus responsible for the AB-fluxes, which will appear in the next section. $\Pic^N(\Sigma)$ carries a natural $(1,1)$-cohomology class $\Theta$, Poincar\'e-dual to the theta-divisor, and the projective spaces carry another cohomology class $\xi$, dual to the divisor of
configurations of~$N$ points where at least one point coincides with a fixed point on~$\Sigma$. The class $\xi$ restricts to the hyperplane class in each fiber~$\mathbb P^{N-\g}$.

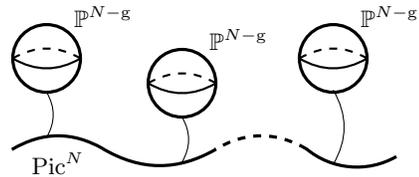
\begin{figure}[h]
\begin{center}
\begin{tikzpicture}[smooth cycle,scale=0.30]
\tikzset{->-/.style={decoration={
  markings,
  mark=at position .5 with {\arrow{>}}},postaction={decorate}}}
  \tikzset{-<-/.style={decoration={
  markings,
  mark=at position .5 with {\arrow{<}}},postaction={decorate}}}
\draw [black,very thick,bend right] (10+9,1) arc (-130:-60:3.5);
\draw [black,very thick,bend right] (10.01,0.99) edge (5.96,.96);
\draw [black,very thick,bend left] (10+5,0.97) edge (6.0+4,.99);
\draw [black,dashed,very thick,bend right] (10.01+9,0.99) edge (5.96+9,.96);
\draw [black,very thick,bend right] (9,5) arc (0:360:1.5);
\draw [black,bend right] (7.5,1.5) edge (7.5,3.5);
\draw [black,thick,bend right] (6,5) edge (9,5);
\draw [black,dashed,thick,bend left] (6,5) edge (9,5);
\draw [black,very thick,bend right] (9+6,5-1.1) arc (0:360:1.5);
\draw [black,bend right] (7.5+6,1.5-1.1) edge (7.5+6,3.5-1.1);
\draw [black,thick,bend right] (6+6,5-1.1) edge (9+6,5-1.1);
\draw [black,dashed,thick,bend left] (6+6,5-1.1) edge (9+6,5-1.1);
\draw [black,very thick,bend right] (9+12.7,5) arc (0:360:1.5);
\draw [black,bend right] (7.5+12.7,.3) edge (7.5+12.7,3.6);
\draw [black,thick,bend right] (6+12.7,5) edge (9+12.7,5);
\draw [black,dashed,thick,bend left] (6+12.7,5) edge (9+12.7,5);
%\draw [black,very thick,bend right] (11+9,3.7) edge (13.8+9,3.7);
%\draw [black,very thick,bend left] (11.3+9,3.55) edge (13.5+9,3.55);
%\draw [red,dashed,very thick,bend right,->-] (5.7,3.8) arc (-20:-160:2.4);
%\draw [green,dashed,very thick,bend right,->-] (5.7+9,3.8) arc (-20:-160:2.4);
%\draw [violet,dashed,very thick,bend right,->-] (5.7+18,3.8) arc (-20:-160:2.4);
%\draw [blue,very thick]  (3.5,3.5) edge [->] (3.5,6);
%%\draw [blue,dashed,very thick,bend right]  (3.5,3.3) edge [-<-] (3.5,0.2);
%\draw [cyan,very thick]  (3.5+9,3.5) edge [->] (3.5+9,6);
%%\draw [cyan,dashed,very thick,bend right]  (3.5+9,3.3) edge [-<-] (3.5+9,0.2);
%\draw [orange,very thick]  (3.5+18,3.5) edge [->] (3.5+18,6);
\draw[black] (8,-.5) node [above]{$\Pic^N$};
\draw[black] (10,6) node [above]{$\mathbb P^{N-\g}$};
\draw[black] (16,5) node [above]{$\mathbb P^{N-\g}$};
\draw[black] (22.7,6) node [above]{$\mathbb P^{N-\g}$};
%\draw[black] (4.6+9,4.6) node [above]{$\phi_2$};
%\draw[black] (4.2+9,.4) node [above]{$\phi_{2+\g}$};
%\draw[black] (4.6+18,4.6) node [above]{$\phi_\g$};
%\draw[black] (4.2+18,.4) node [above]{$\phi_{2\g}$};
% %       \draw[black] (8.15,3.55) node [above]{$P_0$};
%  %              \fill [black] (8,3.6) circle (3pt);
\end{tikzpicture}
{\small \caption{Representation of the $N$th symmetric power of the Riemann surface as projective spaces $\mathbb P^{N-\g}$ fibered over the Picard variety.}
\label{fig:jacobian}}
\end{center}
\end{figure}

We adopt the notation $\Theta_{\rm conf}$ for the class $\Theta$ in $\Pic^{N}(\Sigma)$, arising from the configurations of $N$ points on the surface, in order to distinguish it from another class $\Theta$ in $\Pic^{N_\phi}(\Sigma)$ on the space of AB-fluxes. 
The first Chern class of the line bundle $S = L^{\boxtimes N}(-\beta\Delta)$ over $X = S^N\Sigma$ then equals 
\begin{equation}\label{c1}
c_1(S)=\beta\Theta_{\rm conf}+p\xi,
\end{equation} 
where $p=N_\phi-\beta(N+\g-1)$, and all three classes $\Theta_{\rm conf}$, $\xi$, and $c_1(S)$ lie in $H^2(X,\mathbb{Z})$. Further, the Todd class of~$X$ reads
\begin{equation} \label{td(X)}
{\rm td} (X)=({\rm td}\,\xi)^{N-\g-1}\exp\left(\Theta_{\rm conf}\frac{{\rm td} \,\xi-1-\xi}{\xi}\right),
\end{equation}
where ${\rm td}\,\xi=\frac{\xi}{1-e^{-\xi}}$. This is a mixed degree even cohomology class spanning all even degrees from~$0$ to~$2\dim_{\mathbb C}X$. Plugging this into Eq.\ \eq{rank1}  we arrive at the following formula for the dimension of the vector space of Laughlin states, $r=r(N,\beta,p,\g)$, and consequently for the rank of the Laughlin bundle:
\begin{equation}\label{rank}
r=\sum_{k=0}^\g
{{\g}\choose{k}}{{N-\g+p}\choose{k-\g+p}}\cdot \beta^k,
\end{equation}
with the convention ${{a}\choose{b}}=0$ if $b<0$. 
%Added this:
Since both $c_1(S)$ and ${\rm td}(X)$ are expressed in terms of $\xi$ and $\Theta_{\rm conf}$ and that the intersection numbers of these two classes are known, deducing the formula for the rank~$r$ is a purely combinatorial problem, but it is not entirely trivial; actually, the computation involves the Lagrange inversion theorem \cite{KZ}.

It follows from~\eqref{rank} that there are no Laughlin states for $p<0$, in other words for the given filling fraction $1/\beta$ and the magnetic flux $N_\phi$, the configuration of $N=N_{\rm max}$ particles, where
\begin{equation}\label{Nmax}
N_{\rm max}=\left[\frac{N_\phi}{\beta}\right]+1-\g,
\end{equation}
is optimally packed in the sense that no extra particles can be added on the lowest Landau level. Moreover, the state with $N=N_{\rm max}$ when $N_\phi$ is divisible by $\beta$, is incompressible, and this results demonstrates the Wen-Zee shift formula Ref.\ \cite{Wen1992a}. In the latter case the degeneracy of the Laughlin states 
\[
r|_{p=0}=\beta^\g 
\]
is purely topological, that is, independent of~$N$. Thus our result establishes the topological degeneracy of Wen-Zee \cite{WenNiu1990} (see Haldane-Rezayi \cite{HaldaneRezayi} for the case of a torus). The explicit expression for the Laughlin states in the optimally packed configurations we refer to Ref.\ \cite{Kl2019}.

For the case $p>0$, formula \eq{rank} computes the dimension of the full many-body Hilbert spaces of Laughlin states, corresponding to sub-optimally packed configurations, generalizing the ${{N+p}\choose{p}}$ degeneracy of Eq.\ \eq{Ls} on the sphere \cite[Eq.\ 3]{Haldane1991} and \cite[Eq.\ 6]{Li1996} on the torus.

Now, the case of $p>0$ quasiholes localized at fixed points $w_1,\ldots,w_p$ corresponds to a subspace of the full many-body Hilbert space analogous to the one in Eq.\ \eq{Lsq}, but also degenerate for $\g>0$. The same calculation goes through in this case with the replacement of the original line bundle $L$ by the line bundle $L(-w_1-\ldots-w_p)$. Thus plugging $N_\phi\to N_\phi-p$ we obtain that the dimension of this subspace again equals~$\beta^\g$.

\vspace{0.3cm}
\noindent{\it 5. Chern classes of the Laughlin bundle} \; Turning on the solenoid AB-fluxes, see Fig.\ \ref{fig:highergenus}, brings in the parameter space $M=\Pic^{N_\phi}(\Sigma)$ of real dimension $2\g$: the number of independent fluxes through cycles on the surface. In this setting the Hall conductance $\sigma_H$ was computed for the IQHE in Ref.\ \cite{Avron1994} as the first Chern class of the Laughlin bundle on~$M$. Here we generalize this result to the FQHE and give an explicit expression for the Hall conductance.

Again we define Laughlin states as sections of a line bundle $S =L^{\boxtimes N}(-\beta\Delta)$, but now over the product space $M \times X$, where once more, $X=S^N\Sigma$ is viewed as a  $\mathbb P^{N-\g}$ bundle over the Picard variety $\Pic^N(\Sigma)$. Since as manifolds both $M = \Pic^{N_\phi}(\Sigma)$ and $\Pic^N(\Sigma)$ are isomorphic to the same $2\g$-dimensional torus, we distinguish them in what follows by putting prime on the object related to the latter.
 
In order to describe how the AB-fluxes couple to the electronic states, we makes use of the canonical basis of 1-cycles on $\Sigma$, which are $\g$ pairs of simple loops $(\gamma_a,\gamma_{\g+a})$ for each handle of the surface. Let $(\alpha_a,\beta_a)$ be the corresponding dual basis of harmonic 1-forms on~$\Sigma$.  Then insertion of AB-fluxes $\phi_a, \phi_b$ leads to the change of the one-particle $U(1)$ electromagnetic connection $\nabla_z\to\nabla_z+\sum_{a=1}^\g(\phi_a\alpha_a+\phi_{a+\g}\beta_a)$, \cite[Eq.\ 3]{Avron1994}. Taking the trivial connection along $\Pic^{N_\phi}(\Sigma)$, $\nabla_\phi=\sum_a( d\phi_a\p_{\phi_a}+d\phi_{a+\g}\p_{\phi_{a+\g}})$ and summing over all $N$ particles, we arrive at the expression for the first Chern class generalizing Eq.\ \eq{c1},
$$
c_1(S)=
\beta\Theta_{\rm conf}+p\xi
+\sum_{a=1}^\g(d\phi_a\wedge d\phi'_{a+\g}+d\phi'_a\wedge d\phi_{a+\g}),
$$
where the novel term is a two-from with one component along $M=\Pic^{N_\phi}(\Sigma)$ and the other, primed, component along $\Pic^N(\Sigma)$. 

Now we consider this problem in the setting of adiabatic transport, with the space of AB-fluxes $\phi_a$ as the parameter space. From this point of view, the Laughlin bundle is a rank-$r$ vector bundle over $M = \Pic^{N_\phi}(\Sigma)$, i.e., a vector bundle whose fibers are $r$-dimensional vector spaces of Laughlin states. In order to compute the Chern classes of this bundle as cohomology classes on  
$M$, we apply the Grothendieck–Riemann–Roch theorem,
\[
{\rm ch} (V)=\int_Xe^{c_1(S)} {\rm td}(X),
\]
and the result of the integration over the fibers $X$ in the product $X \times M$ is the Chern character of this vector bundle on~$M$. 
We recall that the Chern character is the sum
\begin{align}\nonumber
{\rm ch}(V)={\rm ch}_0(V)+{\rm ch}_1(V)+{\rm ch}_2(V)+\ldots,
\end{align}
of pure degree cohomology classes ${\rm ch}_m(V)$, where ${\rm ch}_0(V)$ is the rank, ${\rm ch}_1(V) = c_1(V)$, and, more generally, the full Chern class is recovered from the Chern characters by the formula
\[
c(V) = \exp \bigg[ \sum_{i\geq1}(-1)^{i-1}(i-1)!\, {\rm ch}_i(V) \bigg].
\]

Performing the integration we obtain the following result
\begin{equation}\label{chm}
{\rm ch}_m(V)=\sum_{k=m}^\g
{{\g-m}\choose{k-m}}{{N-\g+p}\choose{k-\g+p}}\beta^{k-m} \frac{\Theta_{\rm flux}^m}{m!}.
\end{equation}
As a consistency check, for $m=0$ we do recover Eq.~\ref{rank} for the rank. The theta class $\Theta_{\rm flux}$ can be represented as 
\[
\Theta_{\rm flux}=\sum_{a=1}^\g d\phi_a\wedge d\phi_{a+\g}.
\]

When the ground state is completely filled, $p=0$, only last term on the rhs of Eq.\ \eq{chm} remains and the total Chern character sums up to
\[
{\rm ch}(V)= r\,e^{\frac{1}{\beta}\Theta_{\rm flux}}, \quad r=\beta^\g.
\]
Hence the first Chern class is given by
\begin{equation}\label{Chcl}
c_1(V)=\beta^{\g-1}\cdot\Theta_{\rm flux}
\end{equation}
and all the higher Chern classes are the powers of the first Chern class 
\[
c_i(V)={{\beta^\g}\choose{ i}}\cdot \left[\frac{\Theta_{\rm flux}}\beta\right]^i,\quad1\leq i\leq \g,
\]
as is consistent with formula \eq{test} for projectively flat bundles.
This formula stands unchanged for the case of $p$ localized quasiholes, where the degeneracy is still $\beta^\g$, as discussed in the end of \S 4. 

However when we have $p$ non-localized quasiholes, the rank of the bundle is given by \eq{rank} and the relation~\eq{test} does not hold. We can conclude that in this case the Laughlin bundle is definitely not projectively flat.

\vspace{0.3cm}
\noindent{\it 6. Hall conductance and projective flatness test.} \; Following Refs.\ \cite{Thouless1985,TaoWu,Avron85} we consider now the charge transport on our surface in the setting of higher genus surface \cite{Avron1994}.
In the case when the Laughlin state is either completely filled \eq{Nmax} or has all $p$
quasiholes completely localized, the topological contribution to the Hall current reads $I_a=\sigma_{a,a+\g}V_{a+\g}$. Thus changing the AB flux through the cycle $a+\g$ of the surface induces the Hall current in the dual cycle $a$ controlled by the precisely quantized Hall conductance 2-form, given by formula Eq.\ \ref{cond},
\[\sigma_H=\frac1\beta\Theta_{\rm flux}.
\]
We stress that this equation remains exact for any number of particles $N$.
Next we increase the magnetic flux $N_\phi$ until we find ourselves in the situation when there are more quasiholes than the number of impurities that can localize them. If there $p$ extra non-localized fluxes of the magnetic field, we are in the setting of the full many-body  
Hilbert space \eq{Ls} whose dimension, i.e. the rank of the corresponding Laughlin bundle, is given by Eq.\ \eq{rank}. Its first Chern class reads
\[
c_1(V)=\sum_{k=1}^\g
{{\g-1}\choose{k-1}}{{N-\g+p}\choose{k-\g+p}}\cdot\beta^{k-1}\Theta_{\rm flux}.
\]
These states no longer pass the projective flatness test 
and one of the immediate most striking consequences is that the Hall current is no longer precisely quantized. Indeed, taking the large $N$ asymptotics of the rank and the first Chern class, while keeping $p$ and $\g$ fixed we arrive at the following asymptotic expression for the Hall conductance 
\[
\sigma_H=\left(\frac1\beta-\frac{p}{\beta^2\g N}+\mathcal O(1/N^2)\right)\Theta_{\rm flux}.
\]
We see that as soon as non-localized quasiholes proliferate the Hall conductance starts to deviate from its precise quantized value $1/\beta$. At the same time the Hall conductance decreases, which is consistent with the fact that increasing $p$ corresponds to increasing the flux of the perpendicular magnetic field. This formula generalizes the  FQHE result, see e.g. \cite[Eq.\ 4.72]{Yoshioka2002}, which corresponds to the $\g=1$ case. 

\vspace{0.3cm}
\noindent
{\it\ 7. Discussion}\;  
By definition, the Hall conductance is the closed 2-form on the parameter space~$M = \Pic^{N_\phi}(\Sigma)$ given by the curvature of the adiabatic Berry connection of the Laughlin bundle. Our computation only determines the cohomology class represented by this 2-form, but not form itself. The adiabatic connection, which preserves the $L^2$ structure, requires a computation of $N$-fold $L^2$-normalization integrals, see e.g.\ \cite{Kl2015} for a review. In the IQHE, i.e. for $\beta=1$, the wave functions are given by the Slater determinant, and the adiabatic curvature can be computed explicitly~\cite{Avron1994,Klevtsov:2017tt}. In this case, the result for the Hall conductance, as $N$ becomes large, is exponentially in $N$ close to the one in \eq{Chcl}, as Avron-Seiler-Zograf have demonstrated using the Quillen metric on determinant line bundles. We expect the same effect to hold for $\beta\neq1$, i.e. the adiabatic curvature being exponentially close to~\eq{Chcl} for large $N$, although this point definitely deserves further investigation.  In the case of FQHE and $\Sigma$ being a torus, the adiabatic connection for the fully-filled ($N_\phi=\beta N$) Laughlin states is known to be projectively flat, see e.g.~\cite{KVW,Varn} and \cite[\S 4.2.]{Kl2015}. We further note that the importance of the projective flatness of the quantum bundles for the consistency of the general quantization procedure was emphasized in Ref.~\cite{Axelrod}.

We have seen that the bundles of quantum states passing the projective flatness test do turn out to be sufficiently robust and thus warrant the label of topological states of matter. 
%However, these states are severely restricted by the condition that their higher Chern classes are given by powers of the first Chern class. In our opinion, there could be another interesting possibility, when the quantum states do possess an independent second, or higher Chern classes and at the same time are sufficiently robust to correspond to an experimentally realized phase of matter. It would be interesting to explore this possibility. 
It would be interesting to apply our test to other FQHE states \cite{Moore1991,ReadRezayi1996,Bernevig2008,Jain,HHSV} and other parameter spaces \cite{Avron1995,Levay,Tokatly2007,KW2015,Read2015,CCLW,Barkeshli2013,Gromov2016}, such as the moduli spaces of complex structures on $\Sigma$ where projective flatness has been conjectured to hold for some of the states in FQHE~\cite{Read2009}. In the case of the parameter spaces being the space of positions of the quasiholes, our test can be applied to the question of topological braiding \cite{Halperin1984,Arovas1984,Bonderson2011}.

We thank Y.~Avron for careful reading and detailed comments on the manuscript and P.~Wiegmann for useful discussions. The work of S.K. was partly supported by the IdEx program and the USIAS Fellowship of the University of Strasbourg, and the ANR-20-CE40-0017 grant. The work of D.Z. was partly supported by the ANR-18-CE40-0009 ENUMGEOM grant.

\bibliography{adiabaticphase_refs}
\end{document}